\documentclass[conference]{IEEEtran}
\IEEEoverridecommandlockouts
\usepackage{cite}
\usepackage{amsmath,amssymb,amsfonts}
\usepackage{subcaption}
\usepackage{caption}
\usepackage{multicol}
\usepackage{longtable}
\usepackage{booktabs}
\usepackage{graphicx}
\usepackage{textcomp}
\usepackage{xcolor}
\usepackage{tikz}
\usepackage[inline]{enumitem}

\newcommand{\roundbox}[1]{
	\begin{center}
		\begin{tikzpicture}
		\node[draw=black, rectangle, rounded corners](box){
			\begin{minipage}{0.95\columnwidth}
			#1
			\end{minipage}
		};
		\end{tikzpicture}
	\end{center}
}

\def\BibTeX{{\rm B\kern-.05em{\sc i\kern-.025em b}\kern-.08em
    T\kern-.1667em\lower.7ex\hbox{E}\kern-.125emX}}

\usepackage[utf8]{inputenc}
\usepackage{listings}
\usepackage{xcolor}
\definecolor{codegreen}{rgb}{0,0.6,0}
\definecolor{codegray}{rgb}{0.5,0.5,0.5}
\definecolor{codepurple}{rgb}{0.58,0,0.82}
\definecolor{backcolour}{rgb}{0.95,0.95,0.92}
\newcommand\footnoteref[1]{\protected@xdef\@thefnmark{\ref{#1}}\@footnotemark}
\lstdefinestyle{mystyle}{
	backgroundcolor=\color{backcolour},   commentstyle=\color{codegreen},
	keywordstyle=\color{magenta},
	numberstyle=\tiny\color{codegray},
	stringstyle=\color{codepurple},
	basicstyle=\ttfamily\footnotesize,
	breakatwhitespace=false,         
	breaklines=true,                 
	captionpos=b,                    
	keepspaces=true,                 
	numbers=left,                    
	numbersep=5pt,                  
	showspaces=false,                
	showstringspaces=false,
	showtabs=false,                  
	tabsize=2
}
\label{firstlist}
\usepackage{lineno,hyperref}
\modulolinenumbers[5]
\begin{document}

\title{An empirical evaluation of the usefulness of Tree Kernels for Commit-time Defect Detection in large software systems\\
}

\author{\IEEEauthorblockN{1\textsuperscript{st} Hareem Sahar}
\IEEEauthorblockA{\textit{Dept. of Computing Sciences} \\
\textit{University of Alberta}\\
Edmonton, Canada\\
}
\and
\IEEEauthorblockN{2\textsuperscript{nd} Yuxin Liu}
\IEEEauthorblockA{\textit{Dept. of Computing Sciences} \\
\textit{University of Alberta}\\
Edmonton, Canada\\
}
\and
\IEEEauthorblockN{3\textsuperscript{rd} Abram Hindle}
\IEEEauthorblockA{\textit{Dept. of Computing Sciences} \\
\textit{University of Alberta}\\
Edmonton, Canada \\
}

\and	
\IEEEauthorblockN{4\textsuperscript{th} Denilson Barbosa}
\IEEEauthorblockA{\textit{Dept. of Computing Sciences} \\
\textit{University of Alberta}\\
Edmonton, Canada  \\
}

}

\maketitle

\begin{abstract}
Defect detection at commit check-in time prevents the introduction of defects into software systems. 
Current defect detection approaches rely on metric-based models which are not very accurate and whose results are not directly useful for developers. 
We propose a method to detect bug-inducing commits by comparing the incoming changes with all past commits in the project, considering both those that introduced defects and those that did not.
Our method considers individual changes in the commit separately, at the method-level granularity.
Doing so helps developers as they are informed of specific methods that need further attention instead of being told that the entire commit is problematic.
Our approach represents source code as abstract syntax trees and uses tree kernels to estimate the similarity of the code with previous commits. 
We experiment with subtree kernels (STK), subset tree kernels (SSTK), or partial tree kernels (PTK). 
An incoming change is then classified using a K-NN classifier on the past changes. 
We evaluate our approach on the BigCloneBench benchmark and on the Technical Debt dataset, using the NiCad clone detector as the baseline. 
Our experiments with the BigCloneBench benchmark show that the tree kernel approach can detect clones with a comparable MAP to that of NiCad. Also, on defect detection with the Technical Debt dataset, tree kernels are least as effective as NiCad with MRR, F-score, and Accuracy of 0.87, 0.80, and 0.82 respectively. 
\end{abstract}


\section{Introduction}
\label{intro}
Defect detection and prevention is an important yet costly maintenance activity. The later a defect is identified in the software development life cycle, the higher its impact and cost. To identify and prevent defects early on, Just-in-time (JIT) defect detection~\cite{mockus2000predicting} is used as it can identify risky changes at commit check-in time. It is also referred to as JIT defect prediction~\cite{hoang2019deepjit}, JIT fault detection~\cite{clever}, JIT defect identification~\cite{yan2020JIT}, and JIT Quality Assurance~\cite{kamei2012large}. The recent work on JIT defect detection was carried out by Nayrolles and Hamou-Lhad~\cite{clever}, and Yan \emph{et al.}~\cite{yan2020JIT}. These are state-of-the-art approaches that use metric-based regression models for defect detection. Metric-based models~\cite{commitguru}\cite{clever}~\cite{yan2020JIT}, often fail to capture the semantic and syntactic structure of source code changes, and are prone to a high number of false positives. Other recent approaches such as DeepJIT~\cite{hoang2019deepjit} and CC2Vec~\cite{hoang2020cc2vec} require training deep neural network models that can often be cost prohibitive.

In this paper, we propose an approach based on tree kernels for JIT defect detection. Although not widely used in Software Engineering, tree kernels~\cite{moschitti2006making}~\cite{treekernels2001} are a prominent method of comparing structured data to find similarities. Our approach exploits the structural representation of source code in addition to its textual representation. We believe that these complementary source code representations~\cite{zhang2019novel} improve the source code similarity detection, and lead to fewer false positives than the aforementioned metric-based models. Unlike many recent approaches, our tree kernel approach works without training a sophisticated deep learning model, and is simple, explainable~\cite{pornprasit2021jitline} as well as deterministic.

Our approach works in two steps. First, we retrieve top 100 source code changes that exhibit textual similarity with an incoming change. We then convert the incoming change and the retrieved changes into abstract syntax trees, and measure structural similarity between them using tree kernels. Of several well-known tree kernel methods, we rely upon \textit{Subtree kernel}~\cite{STK}, \textit{Subset tree kernel}~\cite{STK}, and \textit{Partial tree kernel}~\cite{PTK}. We use changes from the past having similarity with an incoming change to classify it.

For evaluation of our approach, we use two datasets, the first one is called BigCloneBench~\cite{svajlenko2015}, a clone detection benchmark containing 8.5 million clones from 43 distinct functionalities. The second is called Technical Debt dataset~\cite{TDD} containing labeled commits from 33 Apache foundation Java projects. Through various performance measures we show that our approach can effectively identify defects. 
The approach correctly classified a new change 82\% of the time by looking at the Top-1 most similar change from the past (1-NN classifier). 
It also shows similar performance to baseline such as NiCad clone detector on identifying similar changes. Our contributions are:
\begin{itemize}
	\item{an approach based on tree kernels to detect bug-inducing changes at commit check-in time;}
	
	\item{an evaluation of three tree kernels: Subset tree kernel, Partial tree kernel and Subtree kernel for the source code similarity detection using BigCloneBench clone detection benchmark;}
	
	\item{an evaluation of the usefulness of tree kernels for defect detection using the Technical Debt dataset~\cite{TDD};}
	
	\item{a comparison of defect detection performance of our approach with the baseline NiCad clone detector using the Technical Debt dataset.}
\end{itemize}

\section{Background and Related Work}
\label{sec:1}
In this section, we review source code representation, code similarity detection, tree kernels, and JIT defect detection.
\paragraph{\textbf{Source Code Representation}}
Source code can be represented as raw text, or at an abstract level in the form of a tree or graph~\cite{zhang2019novel}, or even as a set of paths~\cite{kovalenko2019pathminer}. In this paper, we use \emph{abstract syntax trees} (AST) which are a tree-based data-structure to represent the syntactic structure of source code~\cite{AST2005understanding}. 
ASTs are at a higher level of abstraction than the actual program. They represent the essential structure of code, while omitting certain syntactic redundancies such as punctuation or parentheses. Due to these properties they provide an intermediate code representation~\cite{zhang2019novel} which is suitable for various tasks such as clone detection or code similarity detection~\cite{baxter1998clone}.

\paragraph{\textbf{Code Similarity Detection}}
Two fragments of code which are identical or highly similar are called clones. Primarily, the clones are divided into four types~\cite{surveyCD}; \textbf{Type-1} are identical code fragments that only differ by whitespaces or comments. \textbf{Type-2} are syntactically identical code fragments that differ by identifier names, data types or literal values. \textbf{Type-3} are fragments that have been modified due to the addition or deletion of statements. \textbf{Type-4} are code fragments that semantically perform the same computation with little syntactic similarity. These are also called semantic clones.
Clone detection techniques leverage the syntactic and structural details of the code along with a source code similarity measure to detect clones~\cite{svajlenko2015}. A reliable source code similarity measure is fundamental to many other software engineering tasks as well, including but not limited to plagiarism detection, code recommendation~\cite{holmes2005}, and bug detection~\cite{bugdetect1}. Early approaches used metrics as a measure of similarity but were not very effective. The approaches that followed employed different intermediate representations, and similarity measures. 
{\textit{Textual}} approaches compare string sequences e.g., comments and identifiers, whereas \textit{lexical} approaches~\cite{sourcerercc}~\cite{cpminer2006} normalize textual differences by abstracting various aspects of source code such as identifiers and variable names.
{\textit{Tree-based}} approaches~\cite{clonedr1999}~\cite{deckard2007} measure structural similarity while ignoring lexical differences through the use of Abstract Syntax Trees.
In this paper we use one publicly available state-of-the-art text based hybrid clone detector NiCad as a baseline.

\paragraph{\textbf{Kernel Methods}}
In this paper, we exploit the potential of \textit{tree kernels}, which is a prominent method for comparing structured data, to determine similarity between code fragments. 
Tree kernels~\cite{vert2002tree}~\cite{STK} measure similarity between a pair of trees by projecting data to a higher dimension; feature space, where actual points need not be known but the dot product has to be known. 
The definition of a new tree kernel for a given problem requires specifying a set of features and a kernel function to compare trees in terms of their features~\cite{corazza2010tree}.

Between two trees \textit{T}\textsubscript{1} and \textit{T}\textsubscript{2}, a kernel \textit{K(T}\textsubscript1, \textit{T}\textsubscript2\textit{)}  can be
represented as an inner product between two vectors:
\begin{equation}
K(T_{1}, T_{2}) = h (T_{1}) \: h (T_{2}) 
\end{equation}

Each tree \textit{T} can be represented as a vector as follows, where \textit{h}\textsubscript{i}\textit{(T)} is the number of occurrences of the \textit{i}\textsuperscript{th} subtree in \textit{T}.

\begin{equation}
\textit{h(T)} = (h_{1}(T), h_{2}(T), . . . , h_{n}(T))
\end{equation} 

\begin{figure*}[ht]
	\begin{subfigure}[t]{.26\textwidth}
		\centering
		\includegraphics[width=\columnwidth]{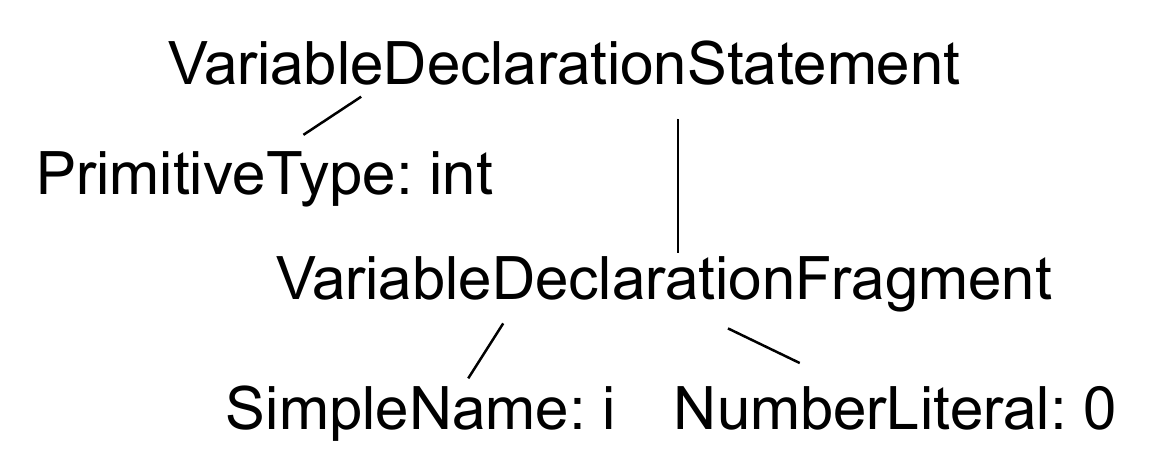} 
		\caption{AST}
	\end{subfigure}%
	\begin{subfigure}[t]{.23\textwidth}
		\centering
		\includegraphics[width=\columnwidth]{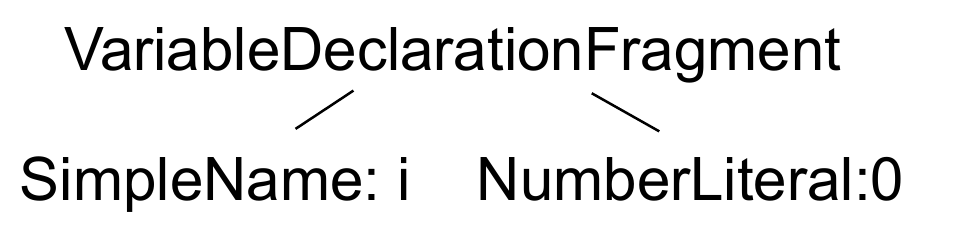}  
		\caption{Subtree Trees}
	\end{subfigure}%
		\begin{subfigure}[t]{.26\textwidth}
		\centering
		\includegraphics[width=\columnwidth]{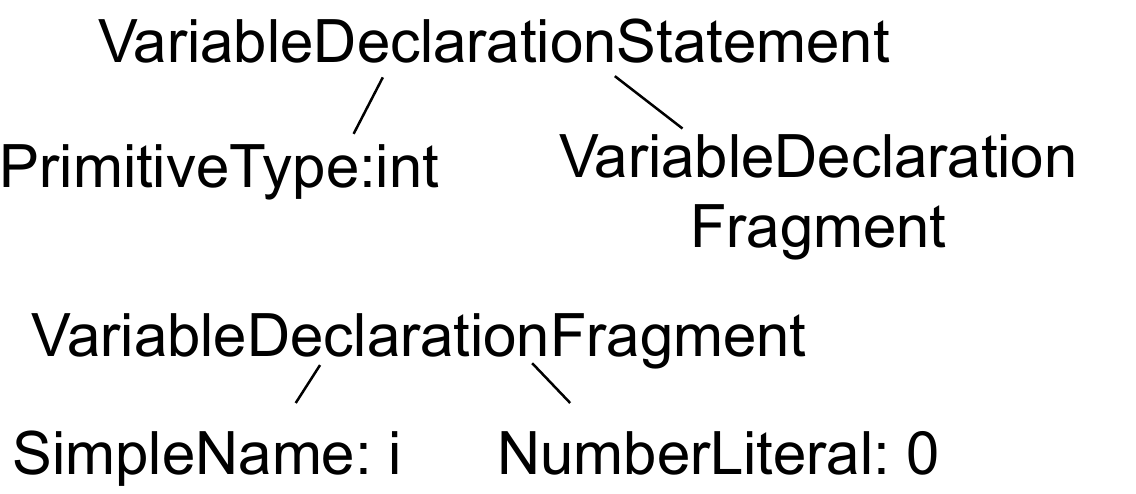}  
		\caption{Subset Trees}
	\end{subfigure}%
	\begin{subfigure}[t]{.25\textwidth}
	\centering
	\includegraphics[width=\columnwidth]{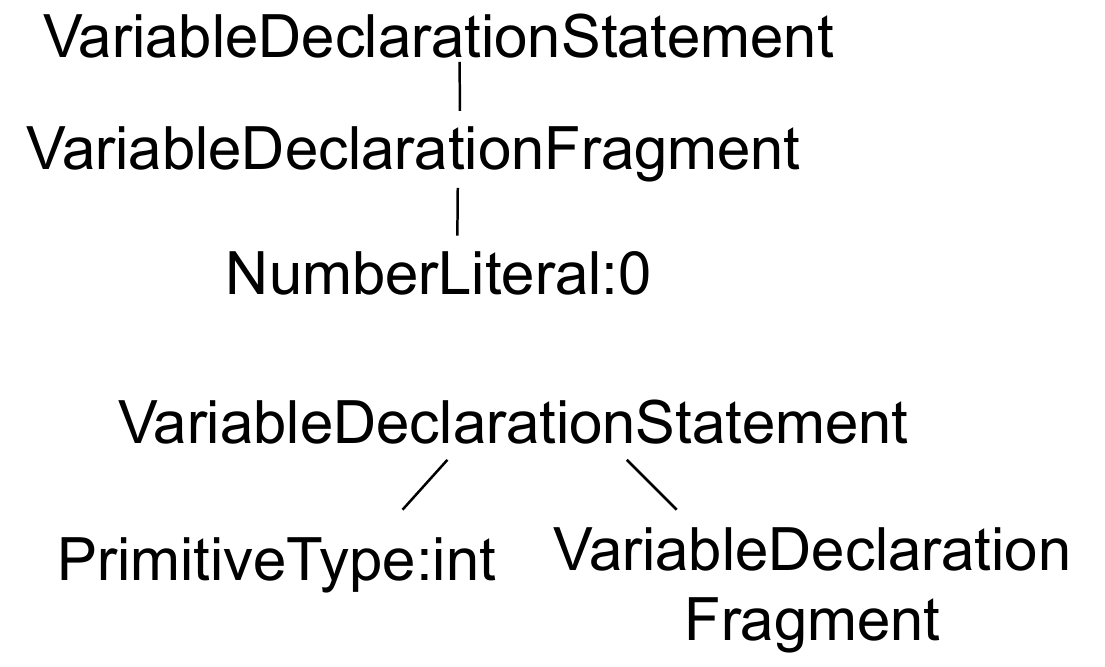}  
	\caption{Partial Trees}
	\label{fig:sub-second}
	\end{subfigure}
	
	\caption{The (a) Abstract syntax tree, (b) some Subtrees, (c) some Subset trees, (d) some Partial trees of the expression int i=0;}
	\label{fig:figA}
\end{figure*}

The \textit{Subtree kernel}~\cite{STKoptimal}~\cite{moschitti2006making} is a convolution kernel that evaluates the number of common subtrees between two given trees, where a subtree is defined as a node and all its children, excluding terminals. In \textit{Subset tree kernel}~\cite{STKoptimal}~\cite{moschitti2006making}, which evaluates the subset-trees, single nodes are not considered as trees. It further assumes that leaves need not be terminals i.e., the leaves can be associated with non-terminal symbols. As a result, there are more subset trees than subtrees for a given sentence. While computing subset trees, a node is either considered with either all of its children or none of them~\cite{filice2017kelp}. This means that the descendancy can be incomplete in depth, but no partial productions are allowed. The \textit{Partial tree kernel}~\cite{PTK} is a convolution kernel that considers partial trees shared between two trees i.e. nodes and their partial descendency
~\cite{filice2017kelp}. Figure~\ref{fig:figA} shows the subtrees, subset trees, and partial trees for the following source code line.

\begin{lstlisting}[caption= A single source code line,label=jsonlisting0]				
										int i = 0; 		
\end{lstlisting}

\paragraph{\textbf{Code Search and Recommendation}}
Code recommenders and code search engines also use similarity detection, whereby an input query is analyzed in its context, and the relevant code snippets are recommended to the developer. Most code search approaches are based on information retrieval (IR). Lu \emph{et al.}~\cite{lu2015} used query expansion with synonyms obtained from WordNet and then performed keyword matching of method signatures whereas others~\cite{Codehow,zhang2017} expanded text query with matched APIs. Gu \emph{et al.}~\cite{deepCS} 
developed a CODEnn model that embeds code snippets and textual descriptions into a unified vector representation, and retrieves code snippets based on those vectors.
Lancer~\cite{lancer} used a Library-Sensitive Language Model (LSLM) and a BERT model to analyze code intent. 
FaCOY~\cite{facoy} finds semantic clones of the query code by employing query alternation and expanding query tokens to include more relevant tokens. Aroma~\cite{aroma} employs structural code search, and  outperforms a TF-IDF approach~\cite{tfidf2002}.

\paragraph{\textbf{Defect Detection and Prevention}}
Defects in source code are commonly detected based on the historical information such as change histories or software quality metrics~\cite{commitguru,clever}. \textit{Traditional defect prediction} approaches relate past defects with source code or process metrics to learn a model that can predict future defects~\cite{buggyorclean}. Static analysis based bug-finders~\cite{findbugs, bugmemories}, on the other hand, use predefined rules inferred from the syntactic and semantic properties of source code to detect coding style bugs such as buffer overflows or infinite recursive loops. 
Due to the repetitive nature of source code~\cite{hindle2012}, statistical language models can also be used to
spot defective code~\cite{ray2016}~\cite{campbell2014syntax}~\cite{santos2018syntax} as effectively as static analysis tools like FindBugs~\cite{findbugs}. 
Interestingly, project specific bugs mined from the version control histories are largely distinct from bugs found by static analysis tools~\cite{bugmemories}, making code-to-code tools complementary to other tools and techniques. 
Another related line of research is \textit{JIT defect detection}, which helps developers produce higher quality code by identifying risky changes just-in-time. CLEVER~\cite{clever} detects new defects in the software by comparing new code to known cases of bug-inducing code. It uses clone detection to improve the detection accuracy of risky changes identified through a metric-based model. DeepBugs\cite{deepbugs} detects incorrect code by reasoning about identifier names, and training a binary classifier on correct and incorrect examples. Yan \emph{et al.}~\cite{yan2020JIT} use several change level features and software naturalness with the N-gram model to localize bug-inducing changes. Their framework ranks bug-inducing changes in order of their suspiciousness. Li \emph{et al.}~\cite{Li2019} proposed an attention-based approach that combines local context and global context using AST paths and program dependency graphs respectively. The approach assigns more weight to the bug-inducing paths and then classifies the code snippet by aggregating different path-based code representations into a single vector. 

\section{Methodology}
\label{sec:methodology}
\begin{figure}
	\centering
	\includegraphics[scale=0.75]{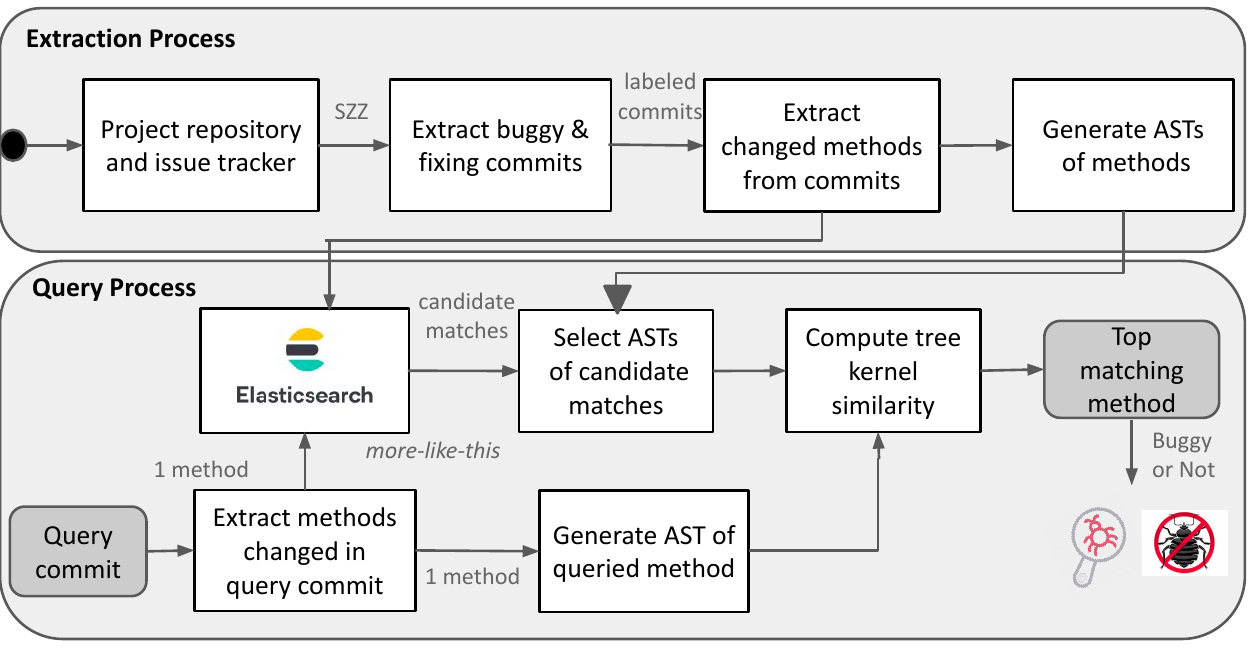}
	\caption{Methodology}
	\label{fig:methodology2}
\end{figure}

In this section, we explain how we collect the past bug-inducing and bug-fixing changes, and later use them to identify defects introduced into a commit during check-in. Figure~\ref{fig:methodology2} provides an overview of our methodology. Our approach begins with extraction of method-level source code changes at commit check-in time. 
The changes are then 
converted into an abstract syntax trees and compared with the past changes from the project, using tree kernels. If a newly committed change matches a bug-inducing change in the existing corpora, it is flagged as risky so the developer can inspect and compare.

\subsection{Extracting bug-inducing and fix-inducing commits}
\label{sec:2a}
As a first step, we build a corpus of bug-inducing and bug-fixing changes. A change corresponds to one or more lines of code that were modified in a commit. In this paper, we work at method-level changes, which may be obtained from the project history using the SZZ algorithm~\cite{szz2005,szz2006}. 
The algorithm works in two steps. In the first step, it identifies the bug-fixing commit using the information available in the version control system and the issue tracking system. The algorithm identifies a commit as bug-fixing if it refers to an issue labeled as fixed in the issue tracking system. In the second step, the algorithm identifies bug-inducing commits using \textit{diff} and \textit{annotate/blame} functionality. The \textit{diff} identifies lines that have been changed between a bug-fixing commit and its previous commit i.e. changes that fix the bug. The \textit{annotate/blame} functionality identifies the commit that modified those lines in the past. SZZ flags the changes as bug-inducing if the commit was made before filing the issue in the issue tracker. Once the SZZ algorithm completes its execution, we have all the bug-inducing and bug-fixing commits for the project.

\subsection{Processing commits to extract methods}
\label{sec:2d}
The source code changes introduced by a commit are spread across one or multiple files. Each changed file has a number of hunks, i.e. a continuous group of lines, that were added, modified, or deleted from a commit~\cite{bugmemories}. 
A hunk may be a partial or a complete block of code and a file can have multiple hunks. Many tree-based structural analysis techniques such as NiCad clone detector (that we use as a baseline) require complete Java code~\cite{nicad2008}. 
Parsing incomplete code is a challenging task itself because of missing context. 
Therefore, we checkout the actual git commit to extract complete method bodies of methods that were modified by a bug-inducing or a bug-fixing commit. These are referred hereafter as modified methods. We then wrap the modified methods in a dummy class definition to facilitate parsing. 

\subsection{Generating Abstract Syntax Trees by parsing methods}
\label{sec:2b}
We transform the modified methods into an abstract syntax tree (AST) using GumTree\footnote{https://github.com/GumTreeDiff/gumtree}. Gumtree~\cite{falleri2014fine} is a source code parsing and \textit{differencing} library that computes fine-grained edit scripts and presents them in a tree format. Here we only use Gumtree to generate ASTs corresponding to each change.
Listing~\ref{jsonlisting1} shows an example code snippet, and its corresponding AST is shown in Listing~\ref{jsonlisting3} in s-expression format. 
The ASTs of all modified methods were stored into an s-expression format for later use.

\begin{lstlisting}[language=Python, caption= An example Java code,label=jsonlisting1]
public class Example {   
	public String foo(int i) { 	
		if (i == 0) return "Foo!";    
	}
}
\end{lstlisting}

\subsection{Classifying incoming changes}
When a developer commits a change, we obtain the complete body of methods that are part of the change. 
The newly extracted method-level changes are then classified in two steps. First, we do a quick search of the existing source code corpus to retrieve any candidate matches. 
Second we do a pairwise comparison between the ASTs of over 100 top search results (i.e., candidate matches) and the newly modified method.
We then classify the new method using a \textbf{K-NN} classifier with K=1 and K=5. The classifier is biased towards buggy changes, i.e., if a bug-inducing change is found in Top-K results, the new change will be classified buggy. Later in our evaluation, we also ensure that each commit is only evaluated against past commits of its project.

\paragraph{Selecting candidate matches}
To make the classification of a new change scalable and efficient, we built an inverted index of the past method-level changes per project. For the purpose of indexing we used Elasticsearch\footnote{https://www.elastic.co/elasticsearch/}, which is a real-time full-text search engine built on top of the Apache Lucene library\footnote{https://lucene.apache.org/}. Each indexed document contains the raw source code of a method that was part of either a bug-inducing or a bug-fixing change. The document also contains meta-data information such as commit hash, file name, method name, and timestamp.
The Elasticsearch index is based on a custom shingle analyzer~\footnote{The min\_shingle\_size=2 and a max\_shingle\_size=3.}, and an edge\_ngram analyzer with standard tokenizer\footnote{The min\_gram=1 and the max\_gram=20.}. For each method-level change extracted from an incoming commit, we query the Elasticsearch index to find the top 100 most similar method-level changes from the indexed corpus. We do so with Elasticsearch's \textit{more\_like\_this} query which utilizes the source code frequencies (from the past) to find changes that match an incoming change. The documents retrieved by Elasticsearch are called candidate matches. Thus by using Elasticsearch we avoid comparing each incoming change with all historical changes, but instead reduce the number of comparisons to only 100 candidates. The choice of using only 100 candidates is made for performance reasons and in future work we will consider other values. 


\paragraph{Computing tree kernel similarity between ASTs}
\label{sec:2c}
Finally, an incoming change is classified by computing its similarity with the candidate matches retrieved from Elasticsearch. However, at this stage, instead of comparing the source code we compare their abstract syntax trees. 
We resort to KeLP's tree kernel implementation for computing similarity between a pair of trees representing an incoming change and a candidate change. KeLP\footnote{https://github.com/SAG-KeLP/kelp-full} is a Java based framework that provides the implementation of kernel functions over strings, trees, and graph~\cite{filice2017kelp}. The output of KeLP is a sorted list of past changes that match an incoming change we are classifying. A label based on K-NN classifier is then assigned to the new change. 

\begin{lstlisting}[ caption= KeLP suitable representation of the abstract syntax tree (AST) generated from Listing~\ref{jsonlisting1} using Gumtree library. This AST is shown in s-expression format that KeLP uses,  label=jsonlisting3]
(CompilationUnit(TypeDeclaration(Modifier:public)(TYPE_DECLARATION_KIND:class)(SimpleName:Example)(MethodDeclaration (Modifier:public)(SimpleType(SimpleName:String))(SimpleName:foo)(SingleVariableDeclaration(PrimitiveType: int)(SimpleName: i))(Block (IfStatement(InfixExpression(SimpleName: i)(INFIX_EXPRESSION_OPERATOR: ==)(NumberLiteral: 0))(ReturnStatement(StringLiteral: "Foo!")))))))
\end{lstlisting}
\section{Evaluation and Results}
\label{sec:eval}
The success of our approach depends on retrieving a code change from the past that looks similar to the newly committed source code change. Therefore, we evaluate our approach in two parts and on two different datasets. In the first part, we determine the performance of tree kernels in detecting similar code (clones), whereas in the second part, we determine the defect detection performance of our approach, using the BigCloneBench benchmark and the Technical Debt dataset~\cite{TDD} respectively. 
Lastly, we compare our results with the results of the NiCad clone detector which is used as a comparison baseline in this work. 

\subsection{Benchmark datasets:}
As mentioned earlier, we carry out two evaluations in this work, first, to evaluate the effectiveness of tree kernels on similar code detection, and second, their effectiveness on defect detection.
\subsubsection{Benchmark for Code Similarity Detection}
Our approach is similar to code similarity or code clone detection in several aspects, hence, we begin by evaluating the performance of tree kernels on a clone detection  benchmark called BigCloneBench~\cite{svajlenko2015}. A clone benchmark contains pairs of code examples which are similar to each other. BigCloneBench has 8.5 million human labeled clones which are mined from the IJadataset 2.0 repository~\cite{ija} that contains 25,000 subject systems spanning 365 MLOC. 
The benchmark clones are divided into 43 distinct functionalities and several clone types. 
Clone pairs that have minor syntactic variations are classified into Type-1 and Type-2, whereas Type-3 and Type-4 are semantic clones. The later two are further divided into sub-categories based on their syntactic similarity values. Very Strongly Type-3 clones range from [0.9, 1.0], Strongly Type-3 from [0.7, 0.9], Moderately Type-3 from [0.5, 0.7], and Weakly Type-3 and 4 range from [0.0, 0.5] on similarity scale. In total, the benchmark has 8,584,153 true positive clones and 279,032 false positive clones constituting 39,951 unique methods altogether. More details about the dataset are available in Svajlenko's dissertation~\cite{thesis2018}. 

\underline{\textit{Evaluation Measures:}}
Our evaluation is based on several binary relevance measures, which are frequently used in prior works~\cite{wen2016locus, yan2020JIT, facoy}. Precision@k computes the relevant responses in the Top-K results for query \textit{q}. It is measured as follows:
\begin{equation}
Precision@k = \frac{1}{|Q|} \sum_{q=1}^{|Q|} \frac{|relevant_{q,k}|}{k}
\label{eq:precEq}
\end{equation}

Mean Average Precision (MAP) is suitable for evaluating an approach that produces multiple responses against each query, ordered by their probability of correctness. For a set of queries, MAP considers the average precision (AP) scores of each query \textit{q} as shown in Eq.~\ref{eq:MAP}.
\begin{equation}
MAP=\sum_{q=1}^{|Q|}\frac{AP_q}{|Q|}
\label{eq:MAP}
\end{equation}

\subsubsection{Benchmark for Defect Detection}
The second part of our evaluation is based on the Technical Debt dataset~\cite{TDD} which contains labeled commits of 33 Java projects from Apache Foundation\footnote{We only use 29 projects that have sufficient bug-inducing and cleans commits for evaluation, see Figure~\ref{fig:bug-inducingfixingpercent}. The excluded projects are commons-beanutils, commons-configuration, commons-daemon, and commons-ognl.}. 
For each project, the dataset contains:
\begin{enumerate*}
 \item{a list of all commits and issues obtained from project's GitHub repository and Jira issue tracker respectively,}\item{a list of refactorings applied to each commit, such as moving or renaming of classes or packages,} \item{a list of code quality related issues per commit, such as code smells, and} \item{a list of changes that induced and fixed bug i.e., bug-inducing and bug-fixing commits.}
\end{enumerate*}

From the dataset, we extracted all changes that added or modified source code within a method in a commit and stored them according to their commit timestamp. We ignored deletions, refactorings, and non-source code changes.
Instead of curating our own data, we relied on an existing benchmark because SZZ implementations are prone to errors~\cite{herbold2019issues}. Using this benchmark helps against potential errors that can affect our study and facilitates replication.

\underline{\textit{Evaluation Measures:}}
For the Technical Debt dataset experiment, we used TopK accuracy, F-score, Accuracy, and Mean Reciprocal Rank. 
TopK accuracy refers to the percentage of changes for which at least one correct result is ranked within the TopK positions in the sorted list of results. We set TopK value for a given change $q$ to $1$, if the change is correctly classified by looking at the top $K$ results only; otherwise, we set the TopK value to $0$. Given a set of $Q$ bug-inducing changes in a project, its TopK accuracy is calculated as follows:

\begin{equation}
TopK_{accuracy} = \frac{1}{|Q|} \sum_{q=1}^{|Q|} TopK(q)
\label{eq:topKEq}
\end{equation}

\emph{F-score} is the harmonic mean of precision and recall where precision is the ratio of TP values to sum of TP and FP values. Recall is ratio of TP values to sum of TP and FN values. 
\begin{equation}
F-score={2\cdot}\frac{{Precision}\cdot{Recall}}{Precision+Recall}
\end{equation}

\emph{Accuracy} is calculated as ratio of sum of TP and TN values to the total number of values.
\begin{equation}
Accuracy=\frac{TP+TN}{TP+FP+TN+FN}
\end{equation}

\emph{Mean Reciprocal Rank} (MRR) is a rank-based measure and is used in evaluations where the user is looking for the best match. For a set of queries, MRR averages the reciprocal rank of the first relevant result returned against a query \textit{q}, where $rank_{q}$ is the rank position of the first relevant match for the query.

\begin{equation}
MRR=\frac{1}{|Q|}{ \sum_{q=1}^{|Q|}} \frac{1}{rank_q}
\label{eq:MRR}
\end{equation}

\subsection{Research Questions}
\textit{RQ1: Can we use a tree kernel based approach to identify inter-project code clones?}

\underline{\textit{{Motivation:}}} The goal of this question is to find out if tree kernels can effectively detect code clones of a query clone in an inter-project setting i.e., code implemented across different real-world projects. The code may be syntactically dissimilar while implementing similar functionality. Our evaluation mimics the real-world settings where a bug-inducing commit resembles a past commit implementing similar functionality.

\begin{figure}
	\includegraphics[scale=0.7]{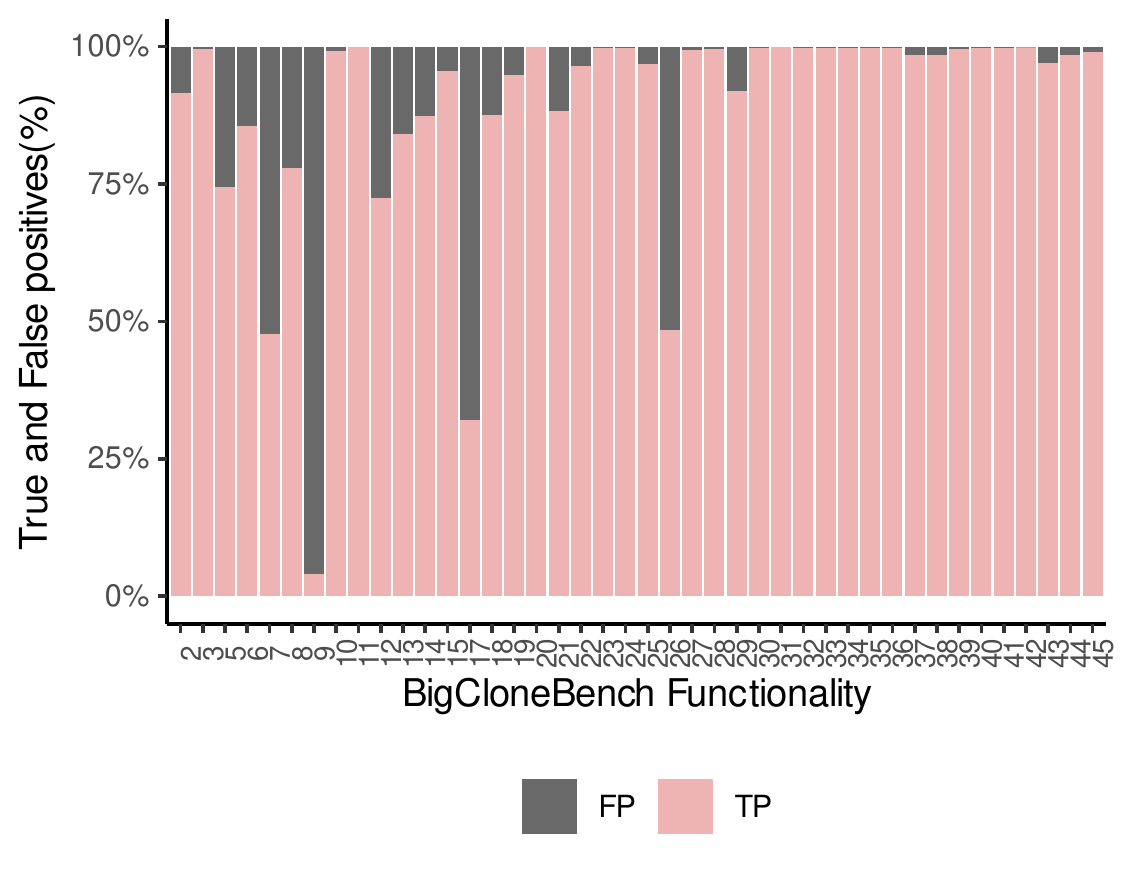}
	\caption{Percentage of true positives (TP) and false positives (FP) per functionality in the BigCloneBench benchmark.}
	\label{fig:truefalse}
\end{figure}

\underline{\textit{{Approach:}}}
To answer this research question, we ran our experiment on BigCloneBench functionalities. The benchmark authors consider individual functionalities, or a random selection of clones from each of the functionalities to be reasonable subsets for achieving high confidence for a clone detector. In our evaluation, we included all the true and false positive clones of a functionality, their breakdown is shown in Figure~\ref{fig:truefalse}. These clones are referred subsequently as candidate clones. 

For each candidate clone \textit{q}, we compute its similarity with the remaining candidate clones using the implementation of tree kernels provided in KeLP library~\cite{filice2017kelp}.  
The matches found by KeLP are ranked based on their similarity with the query clone \textit{q}. The ranked list may also contain false positives. To confirm if a detected match is a true or a false positive, the matches found between candidate clones are compared against the ground truth which is the reference clones in the BigCloneBench benchmark. To determine the number of candidate clones which are correctly matched by KeLP i.e., they are true clones of a given query, we measure Precision@k at K=10.


\underline{\textit{{Results:}}}
Table~\ref{table:map_prec} shows Precision@k and the MAP per BigCloneBench functionality using the three tree kernel methods. The value of \textit{K} is 10, which means to calculate Precision we consider top-10 results, where they exist. The values indicate that tree kernel based approach achieves a good precision for most functionalities. The median values are mostly high, with a slightly lower mean. There is no clear winner among the three tree kernels though. In almost half of the cases PTK performs better than the other two tree kernels, and for the remaining SSTK performs better. STK generally performs worse than the other two.

The MAP values range from 0.02 for \emph{functionality 17} to 0.99 for \emph{functionality 20}. 
If a relevant clone never gets retrieved, we assume the precision corresponding to that to be $0$. 
The MAP for \textit{Fibonacci} is highest, because, for this functionality, there are no false positives in the actual data. One may also observe that functionalities that have a smaller number of false positives, have a higher MAP compared to functionalities that have many false positive candidate clones e.g., \emph{functionality 9}. The reason is that if a given clone has a high similarity with many false positive clones, the true positive clones might be pushed farther in the list, leading to a lower MAP. An ideal similarity detection algorithm should rank the false positives lower than the true positives. Since MAP gives more weight to errors\footnote{error here means an incorrect identification of a clone} that happen high up in the lists, therefore, such an algorithm will achieve high MAP values. Obtaining a higher MAP perfectly aligns with our goal as developers trying to find bugs in their commits would like to see the similar bug-inducing commits (from the past) high up in the results. The MAP per functionality is generally high for many BigCloneBench functionalities leading to median that is higher than NiCad (see Table~\ref{table:map-inter}). The results show that tree-kernel based similarity detection was usually able to retrieve relevant clones against candidate queries. For NiCad, we believe using default configuration was not very effective, and future work should explore other configurations as well.

\roundbox{With a median Precision@10 of 80\% and a MAP of 0.58, our approach can identify clones for a diverse set of functionalities implemented across projects.}

\begin{table}[ht]
	\centering
	\caption{Precision@10 and MAP per BigCloneBench functionality calculated using STK, SSTK, and PTK }
	\label{table:map_prec}
	\begin{tabular}{l||rrr||rrr}
		\toprule
		\multicolumn{1}{c}{} &
		\multicolumn{3}{c}{\textbf{Precision@10}}&
		\multicolumn{3}{c}{\textbf{MAP}}\\
		
		\cmidrule(lr){2-4}
		\cmidrule(lr){5-7}
		
		Funct. & STK & PTK & SSTK & STK & PTK & SSTK \\ 
		\cmidrule(lr){1-1}
		\cmidrule(lr){2-4}
		\cmidrule(lr){5-7}
		
	  2 & 0.10 & 0.20 & 0.20 & 0.11 & 0.13 & 0.12 \\ 
	  3 & 0.30 & 0.60 & 0.50 & 0.25 & 0.29 & 0.28 \\ 
	  5 & 0.60 & 0.60 & 0.60 & 0.38 & 0.38 & 0.38 \\ 
	  6 & 0.80 & 0.90 & 0.80 & 0.62 & 0.64 & 0.62 \\ 
	  7 & 0.20 & 0.80 & 0.40 & 0.20 & 0.30 & 0.23 \\ 
	  8 & 0.50 & 0.35 & 0.50 & 0.15 & 0.14 & 0.16 \\ 
	  9 & 0.30 & 0.30 & 0.40 & 0.12 & 0.13 & 0.13 \\ 
	 10 & 0.80 & 0.80 & 0.80 & 0.58 & 0.59 & 0.59 \\ 
	 11 & 1.00 & 1.00 & 1.00 & 0.93 & 0.93 & 0.93 \\ 
	 12 & 0.60 & 0.80 & 0.80 & 0.34 & 0.44 & 0.41 \\ 
	 13 & 0.40 & 0.60 & 0.50 & 0.24 & 0.31 & 0.28 \\ 
	 14 & 0.00 & 0.50 & 0.20 & 0.15 & 0.23 & 0.18 \\ 
	 15 & 0.80 & 0.80 & 0.80 & 0.64 & 0.71 & 0.65 \\ 
	 17 & 0.10 & 0.00 & 0.10 & 0.02 & 0.03 & 0.02 \\ 
	 18 & 0.30 & 0.50 & 0.40 & 0.36 & 0.40 & 0.38 \\ 
	 19 & 0.20 & 0.40 & 0.30 & 0.26 & 0.30 & 0.27 \\ 
	 20 & 1.00 & 1.00 & 1.00 & 0.99 & 0.99 & 0.99 \\ 
	 21 & 0.50 & 0.60 & 0.70 & 0.40 & 0.43 & 0.45 \\ 
	 22 & 0.30 & 0.40 & 0.40 & 0.18 & 0.21 & 0.19 \\ 
	 23 & 0.80 & 1.00 & 0.80 & 0.79 & 0.85 & 0.80 \\ 
	 24 & 0.80 & 0.90 & 0.90 & 0.76 & 0.78 & 0.76 \\ 
	 25 & 0.40 & 0.40 & 0.40 & 0.23 & 0.25 & 0.24 \\ 
	 26 & 0.00 & 0.30 & 0.00 & 0.03 & 0.07 & 0.03 \\ 
	 27 & 0.70 & 1.00 & 1.00 & 0.55 & 0.62 & 0.60 \\ 
	 28 & 0.40 & 0.90 & 0.80 & 0.52 & 0.58 & 0.56 \\ 
	 29 & 0.20 & 0.30 & 0.20 & 0.24 & 0.25 & 0.24 \\ 
	 30 & 0.80 & 0.90 & 0.00 & 0.58 & 0.61 & 0.59 \\ 
	 31 & 1.00 & 1.00 & 1.00 & 0.94 & 0.94 & 0.93 \\ 
	 32 & 1.00 & 1.00 & 1.00 & 0.87 & 0.88 & 0.87 \\ 
	 33 & 0.90 & 0.90 & 0.90 & 0.83 & 0.83 & 0.84 \\ 
	 34 & 1.00 & 1.00 & 1.00 & 0.86 & 0.88 & 0.87 \\ 
	 35 & 1.00 & 1.00 & 1.00 & 0.90 & 0.91 & 0.90 \\ 
	 36 & 1.00 & 1.00 & 1.00 & 0.92 & 0.91 & 0.91 \\ 
	 37 & 0.10 & 0.40 & 0.10 & 0.17 & 0.23 & 0.19 \\ 
	 38 & 1.00 & 0.90 & 1.00 & 0.84 & 0.83 & 0.85 \\ 
	 39 & 0.30 & 0.80 & 0.50 & 0.46 & 0.55 & 0.50 \\ 
	 40 & 0.80 & 1.00 & 0.90 & 0.68 & 0.73 & 0.70 \\ 
	 41 & 0.80 & 1.00 & 0.90 & 0.83 & 0.87 & 0.84 \\ 
	 42 & 1.00 & 1.00 & 1.00 & 0.88 & 0.88 & 0.88 \\ 
	 43 & 0.30 & 0.80 & 0.30 & 0.32 & 0.44 & 0.37 \\ 
	 44 & 1.00 & 1.00 & 1.00 & 0.95 & 0.95 & 0.95 \\ 
	 45 & 0.70 & 0.80 & 0.80 & 0.55 & 0.58 & 0.58 \\ 
		\bottomrule
	\end{tabular}
\end{table}

\begin{table}[!ht]
	\centering
	\caption{5-statistic summary of the MAP for inter-project clone detection using PTK, STK, SSTK and NiCad} 
	\label{table:map-inter}   
	\centering  
	\begin{tabular}{lrrrrrrr}
		\toprule
		& \textbf{Min} & \textbf{Q$_{1}$} & \textbf{Median} & \textbf{Mean} &  \textbf{Q$_{3}$} & \textbf{Max}\\
		\cmidrule(lr){2-2}
		\cmidrule(lr){3-3}
		\cmidrule(lr){4-4}	
		\cmidrule(lr){5-5}
		\cmidrule(lr){6-6}
		\cmidrule(lr){7-7}
		PTK   &	0.03 & 0.29 & 0.58 &  0.55    & 0.84 & 0.99\\
		STK   &	0.02 & 0.23 & 0.54 &  0.51    & 0.83 & 0.99 \\
		SSTK  & 0.02 & 0.25 & 0.57 &  0.53    & 0.84 & 0.99\\
		NiCad & 0.01 & 0.04 & 0.18 &  0.15    & 0.19 & 0.88\\
		
		\bottomrule
	\end{tabular}
\end{table}


\textit{RQ2: What is the effectiveness of tree kernels in finding intra-project clones?} 

\underline{\textit{{Motivation:}}} The goal of this question is to find out if tree kernels can effectively detect clones of a query in an intra-project setting i.e., within the same subject system. This evaluation mimics the settings where a bug-inducing commit resembles a commit from the past of the project.

\underline{\textit{{Approach:}}}
To answer this question, we ran an experiment using 1000 subject systems from the IJadataset 2.0 whose clones are documented in BigCloneBench. The subject systems were all Java projects implementing multiple common Java functionalities, and constituting thousands of lines of code. The project with the highest number of functions is \textit{jikesrvm}, and the one with the lowest number of function is \textit{nitro-go}. They have 129,601 and 4,182 functions respectively from several BigCloneBench functionalities. Out of 25,000 projects whose clones exist in the BigCloneBench benchmark, we selected the top 1000 projects having highest number of functions.

The candidate clones of each of the 1000 projects were extracted, and for each query clone we find its matches from the other functions that belong to same subject system. This leads to an $O(n^{2})$ time complexity which prohibits scalability~\cite{sourcerercc}. However, for a commit containing a few methods, this comparison was done within seconds.


\underline{\textit{{Results:}}}
The range of MAP values obtained from our intra-project evaluation is shown in Table~\ref{table:map-intra}. These values are calculated across the 1000 projects. All tree kernels perform equally well in intra-project clone detection. The lack of variation in performance seems to be a consequence of each tree kernel having the ability to successfully detect clones in intra-project settings. The results suggest that for 75\% of the projects, relevant clones were returned among the top few results. The high MAP score could be due to the  diverse functionalities implemented in a project. In other words, different implementations of the same function within a project are rare making it easy for tree kernels to identify the  match.

\begin{table}
	\centering
	\caption{MAP summary for intra-project clone detection using PTK, STK, and SSTK and 1000 projects.} 
	\label{table:map-intra}  
	\centering  
	\begin{tabular}{lrrrrrrr}
		\toprule
		& \textbf{Min} & \textbf{Q$_{1}$} & \textbf{Median} & \textbf{Mean} &  \textbf{Q$_{3}$} & \textbf{Max}\\
		\cmidrule(lr){2-2}
		\cmidrule(lr){3-3}
		\cmidrule(lr){4-4}	
		\cmidrule(lr){5-5}
		\cmidrule(lr){6-6}
		\cmidrule(lr){7-7}
		PTK &	0.54 & 0.90 & 1.00 &  0.94 & 1.00 & 1.00\\
		STK &	0.54 & 0.90 & 1.00 &  0.94 & 1.00 & 1.00\\
		SSTK &  0.54 & 0.90 & 1.00 &  0.94 & 1.00 & 1.00\\
		\bottomrule
	\end{tabular}
\end{table}
\roundbox{With a median MAP of 1.0, intra-project clone detection using tree kernels outperforms inter-project clone detection.}

\textit{RQ3: What is the effectiveness of tree kernels in finding clones of different types?}

\underline{\textit{{Motivation:}}} 
In this question we focus on assessing which types of syntactic and semantic clones our approach can effectively identify in BigCloneBench benchmark. 

\underline{\textit{{Approach:}}}
Following Sajnani \emph{et al.}~\cite{sourcerercc}, we only consider Type-1, Type-2, VST3, and ST3 clones in this research question. Using a standard minimum clone size of 6 lines~\cite{svajlenko2014, bellon2007}, we obtained 3,550 Type-1, 993 Type-2, 2,134 VST3, and 4,429 ST3 clones split across several functionalities. 

\underline{\textit{{Results:}}}
For each clone type, the results in terms of mean average precision are shown in Table~\ref{table:typewise}. PTK shows the highest MAP overall followed by SSTK, whereas STK performs poorly. For Type-1 clones SSTK takes the lead by achieving a perfect MAP of 0.98. As expected, the MAP for Type-1 clones is the highest because they are syntactically very similar. This is also the case for NiCad that which has 0.95. Next in order is Type-2 clones with a decent MAP of 0.69, followed by VST3 having a MAP of 0.62. The MAP on ST3 clones is lowest due to their limited syntactic similarity. Overall, \textit{Partial tree kernel} and \textit{Subset tree kernel} perform better than \textit{Subtree kernel} which could not consistently detect clones.
\begin{table}
	\centering
	\caption{MAP per clone type using PTK, STK, SSTK tree kernels, and NiCad clone detector}
	\label{table:typewise}  
	\centering  
	\begin{tabular}{lrrrr}
		\toprule
		&\textbf{Type-1} & \textbf{Type-2} & \textbf{VST3} & \textbf{ST3} \\
		\cmidrule(lr){2-2}
		\cmidrule(lr){3-3}
		\cmidrule(lr){4-4}	
		\cmidrule(lr){5-5}
			
	    	PTK   &  0.91  & 0.69  & 0.62  &  0.50 \\
		    STK   &  0.15  & 0.14  & 0.06  &  0.09 \\
		    SSTK  &  0.98  & 0.67  & 0.60  &  0.42\\
		    NiCad &	0.95   & 0.81  & 0.65  &  0.53    \\
		    
		\bottomrule
	\end{tabular}
\end{table}

\roundbox{PTK and SSTK perform excellent ($>$ 0.9) on Type-1 and good on Type-2 and VST3 clones, but none of the tree kernels perform well on ST3 clones.}

\textit{RQ4: Can we use a tree kernel based approach to detect bug-inducing source code changes in a software system?}

\underline{\textit{{Motivation:}}} Having seen that tree kernels could detect code clones from the BigCloneBench, we aim to find out if a tree kernels approach can correctly classify an incoming source code change into either bug-inducing or fixing. Identifying bug-inducing changes at commit check-in time allows for timely fixing of the buggy code. 

\begin{figure}
	\centering
	\includegraphics[scale=0.8]{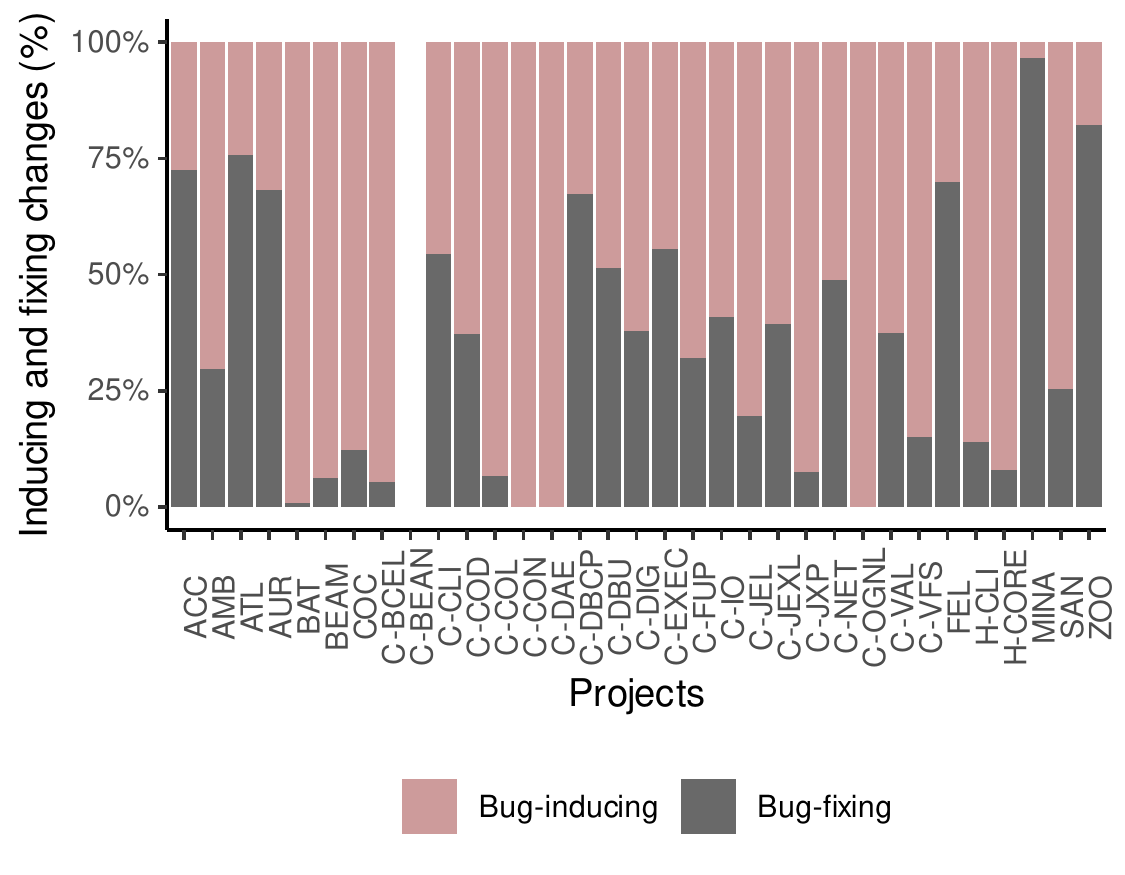}
	\caption{Percentage of actual bug-inducing and bug-fixing changes in the Technical Debt dataset. These changes were identified by the benchmark authors using the SZZ algorithm.}
	\label{fig:bug-inducingfixingpercent}
\end{figure}

\begin{figure}
	\centering
	\includegraphics[scale=0.6]{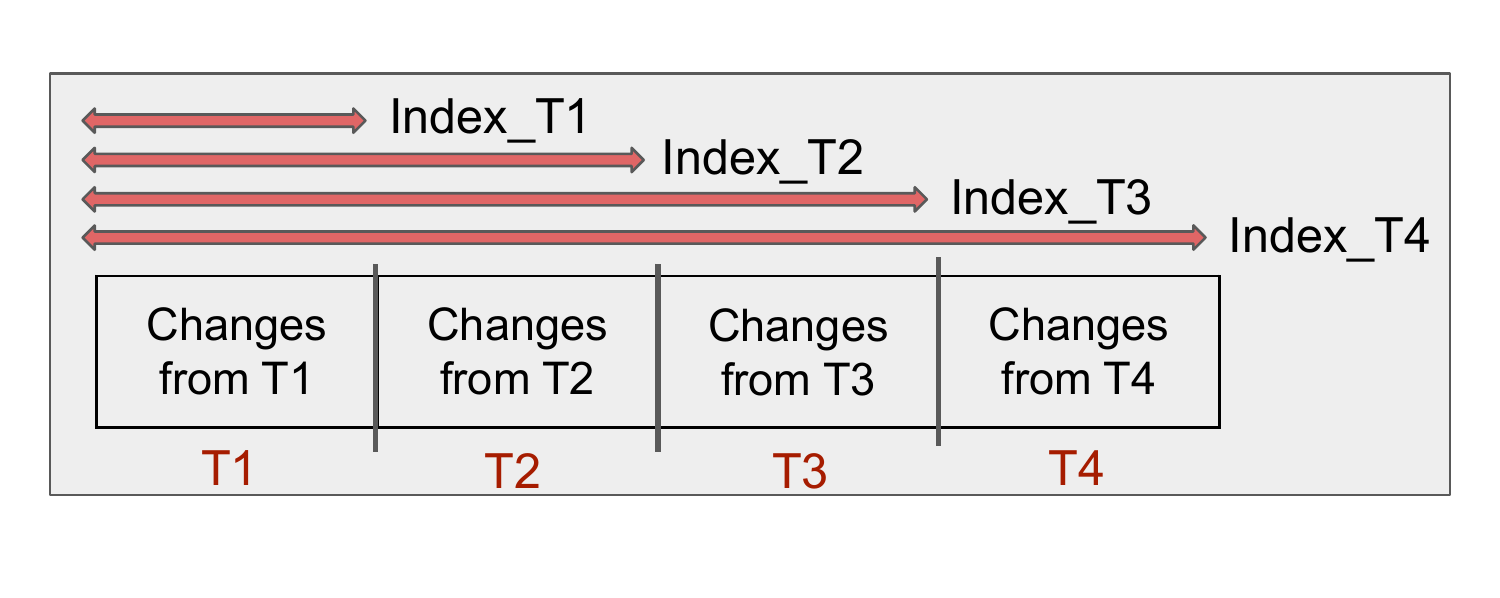}
	\caption{Four Elasticsearch indexes built from changes spread across four time periods. Index\_T1 contains changes from T1, Index\_T2 contains changes from T1 and T2, Index\_T3 contains changes from T1 to T3, and Index\_T4 contains changes from T1 to T4}
	\label{fig:timeaware}
\end{figure}

\underline{\textit{{Approach:}}}
For the evaluation of defect detection performance of our approach, we use both method-level bug-inducing and bug-fixing changes available in the Technical Debt dataset~\cite{TDD}.
We removed the refactoring changes and non-source code changes, as a result some projects did not have enough changes to be evaluated, so we removed them. Figure~\ref{fig:bug-inducingfixingpercent} shows the ratio of remaining bug-inducing and bug-fixing changes. Our assumption is that if a commit has newly added or modified source code, and that commit is identified as bug-inducing, the newly added or modified code was responsible for inducing the bug. Correspondingly, a new source code addition or modification is considered to have resolved the bug in a bug-fixing commit. \\
\textit{\textbf{Time-aware evaluation:}} In practice, an incoming change can only be evaluated against the past project changes, therefore, we evaluate the defect detection approach using a time-aware evaluation~\cite{bangash2020time}. %
We built \textit{N} month-wise Elasticsearch indexes by incrementally adding changes in the order in which they were committed to GitHub. Each index represent one time period,  which was set to 1 month for this study.
Figure~\ref{fig:timeaware} shows changes from 4 time periods. The choice of 1 month long time period is arbitrary and produces a reasonable number of indexes. All the changes from a particular month go together into one index, and each subsequent index contains changes from the next time period, in addition to the past time periods.
Now an incoming change from a particular time period will only be evaluated against past changes from that time period. Such an evaluation does not involve time-travel and ensures realistic performance estimates. 

\underline{\textit{{Results:}}}
The results presented in Table~\ref{table:TDD_nicad_kelp} show that performance varies across projects. 
Tree Kernels achieve a mean MRR, F-score, and accuracy of 0.87, 0.80, and 0.82 respectively. The MRR value signifies that our approach can correctly identify at least 8 changes out of every 10 changes. The mean \textit{TopK@1} value is 0.82 which implies that the first result is correct about 82\% of the times. The mean \textit{TopK@5} is even higher i.e., 0.93. 
This means that a developer only needs to inspect top result in most cases. Occasionally, one might need to inspect the top 5 results to find out a change resembling any incoming change. Since the changes are methods modified or added to the source code files, the actual number of lines that a developer needs to manually inspect is small.
\roundbox{Our approach correctly classifies a new change 82\% of the time using the Top-1 similar change from the project's past.}

\textit{RQ5: Does our tree kernels based approach outperform a clone detection based approach?}

\underline{\textit{{Motivation:}}} Tree Kernels perform well at predicting bugs in newly committed changes. Since our approach resembles code clone detectors in many ways, we compare it against NiCad clone detector, which is known for its effectiveness on clone detection.

\underline{\textit{{Approach:}}}
This experiment aims at evaluating NiCad clone detector (instead of tree kernels) on the Technical Debt dataset .
We use NiCad as baseline because its a well-known clone detector, and prior work CLEVER~\cite{clever} achieved a good performance by using NiCad for validating risky changes. The dataset and experimental design of RQ5 is same as RQ4, except that we replace KeLP similarity computation component with NiCad clone detector version 6.2. Since the RQ4 dataset constitutes method-level changes, it is easy for NiCad to identify clones. Of the clones identified by NiCad, the one that has highest similarity with the query is used for classification. 

\begin{figure}[h]
	\centering
	\includegraphics[scale=0.6]{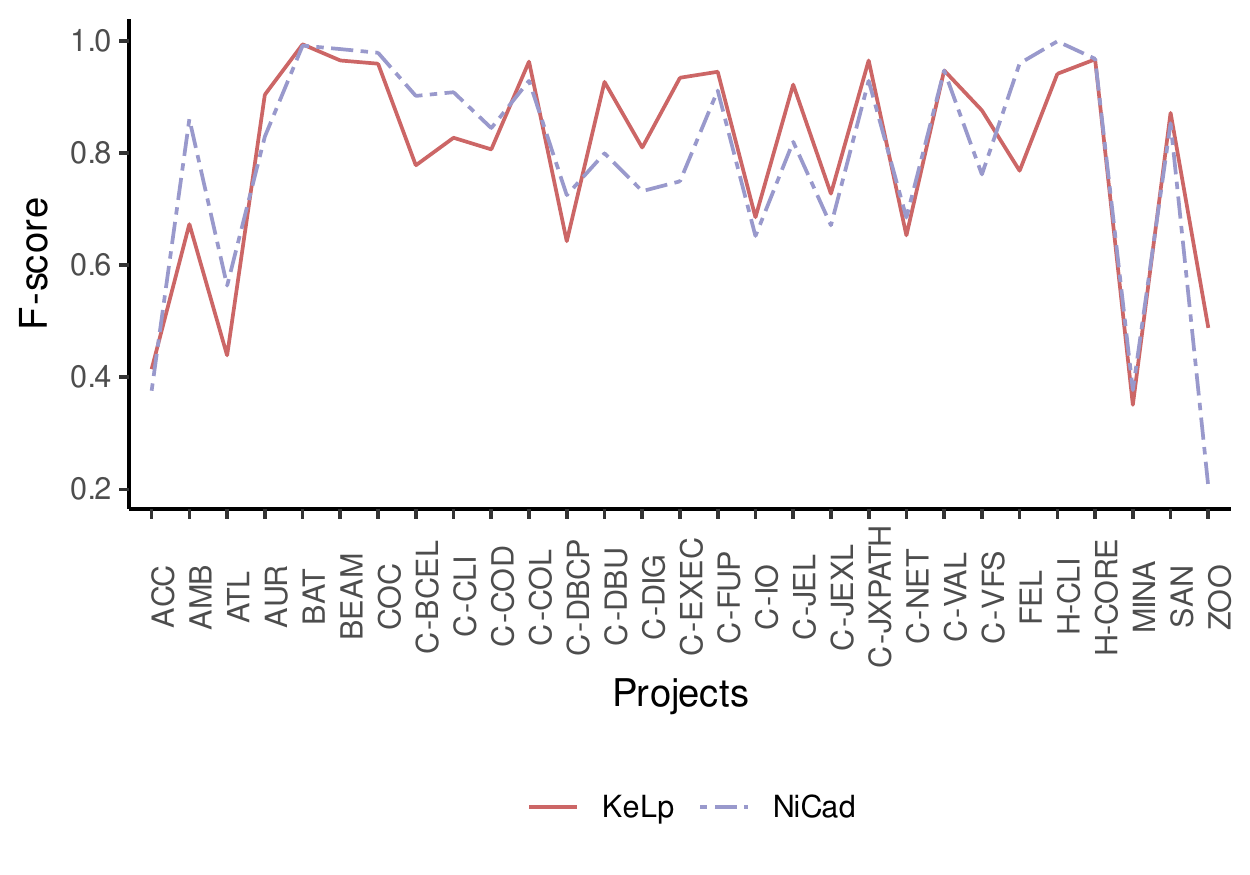}
	\caption{F-score comparison of tree kernels approach with baseline clone detection approach using Technical Debt dataset}
	\label{fig:fscore}
\end{figure}

\underline{\textit{{Results:}}}
The results of baseline are presented in Table~\ref{table:TDD_nicad_kelp}. The baseline approach achieves a mean MRR, F-score, and Accuracy of 0.86, 0.80, and 0.84 respectively. The mean \textit{TopK@1} value is 0.83, which is 0.01 slightly higher than our tree kernel approach, whereas the mean \textit{TopK@5} is 0.90, a drop of 0.03 against our approach. 
The tree kernels are competitive with clone detectors. 
Infact, the rationale behind choosing NiCad as baseline is its exceptional ability to detect clones of different types which gives it an edge over other clone detectors. 
Furthermore, it is easy to run NiCad on the BigCloneBenchmark which makes experimentation easier. 
We used the default dissimilarity threshold of 30\% for NiCad. 
Prior work (CLEVER~\cite{clever}) has also used the same settings to compare code blocks extracted from risky commits. 
Figure~\ref{fig:fscore} compares the F-scores of our approach with the baseline.
Except for one occasion where the baseline F-score and MRR drops to a very low value, the trends are similar across projects. The F-score of both approaches is close with one outperforming the other for some projects.

 Table~\ref{table:TDD_nicad_kelp}, also presents the percentage F-score improvement over the baseline approach. 
A positive value in the \textit{Improved} column indicates our approach outperforms baseline and vice versa. For F-score and MRR, we observed an mean percentage improvement of $5\%$ and $1.8\%$ respectively across all projects using tree kernels. For the accuracy metric, although the baseline outperforms our approach for several projects but the difference is often small to negligible. The mean improvement for accuracy is -2.3\%, i.e., NiCad performs better.
\roundbox{NiCad clone detector correctly classifies a new change 83\% of the times using the Top-1 most similar change from the project's past. The MRR, F-score, and Accuracy metrics also suggest that tree-kernel based approach and NiCad have similar performance on predicting new changes, and neither approach outperforms the other.}

\begin{table*}[ht]
	\centering
		\caption{Defect Prediction performance of STK tree kernel compared against baseline NiCad clone detector on Technical Debt dataset using TopK accuracy (K=1 and K=5), Mean Reciprocal Rank (MRR), F-score, and Accuracy metric. }
		\label{table:TDD_nicad_kelp}
	\begin{tabular}{l||rr|rrr||rr|rrr||r}
		\toprule
	 
		\multicolumn{1}{c}{} &
		\multicolumn{5}{c}{\textbf{KeLP}}&
		\multicolumn{5}{c}{\textbf{NiCad}}&
		\multicolumn{1}{c}{}\\
			
		\cmidrule(lr){3-5}
		\cmidrule(lr){8-10}
		
		\textbf{Project} & \textbf{K=1} & \textbf{K=5} & \textbf{MRR} & \textbf{Accuracy} & \textbf{F-score} &  \textbf{K=1} & \textbf{K=5} & \textbf{MRR} & \textbf{Accuracy} & \textbf{F-score} & \textbf{Improved} \\
		
		\cmidrule(lr){1-1}
		\cmidrule(lr){2-3}
		\cmidrule(lr){4-4}
		\cmidrule(lr){5-5}
		\cmidrule(lr){6-6}	
		\cmidrule(lr){7-8}
		\cmidrule(lr){9-9}
		\cmidrule(lr){10-10}
		\cmidrule(lr){11-11}
		\cmidrule(lr){12-12}
		
		accumulo & 0.58 & 0.94 & 0.73 & 0.58 & 0.41 & 0.62 & 0.76 & 0.68 & 0.62 & 0.38 & 10.2 \\ 
		ambari & 0.60 & 0.79 & 0.69 & 0.60 & 0.67 & 0.78 & 0.90 & 0.83 & 0.78 & 0.86 & -21.8\\ 
		atlas & 0.68 & 0.88 & 0.77 & 0.68 & 0.44 & 0.50 & 0.61 & 0.55 & 0.74 & 0.56 & -22.1\\ 
		attic-aurora & 0.94 & 0.98 & 0.96 & 0.94 & 0.90 & 0.82 & 0.88 & 0.85 & 0.82 & 0.83 & 9.0 \\ 
		 batik & 0.99 & 0.99 & 0.99 & 0.99 & 0.99 & 0.98 & 0.99 & 0.99 & 0.98 & 0.99 & 0.2\\ 
		 beam & 0.94 & 0.98 & 0.95 & 0.94 & 0.97 & 0.97 & 0.97 & 0.97 & 0.97 & 0.99 & -2.1  \\ 
		 cocoon & 0.93 & 0.96 & 0.94 & 0.93 & 0.96 & 0.97 & 0.97 & 0.97 & 0.97 & 0.98 & -2.0 \\ 
		 commons-bcel & 0.64 & 0.94 & 0.77 & 0.64 & 0.78 & 0.83 & 0.85 & 0.84 & 0.83 & 0.90 & -13.7 \\ 
		 commons-cli & 0.81 & 0.97 & 0.89 & 0.81 & 0.83 & 0.95 & 0.96 & 0.95 & 0.95 & 0.91 & -8.9 \\ 
		 commons-codec & 0.75 & 0.89 & 0.82 & 0.75 & 0.81 & 0.82 & 0.90 & 0.86 & 0.82 & 0.85 & -4.5\\ 
		 commons-collections & 0.93 & 0.97 & 0.95 & 0.93 & 0.96 & 0.88 & 0.92 & 0.90 & 0.88 & 0.93 & 3.7 \\ 
		 commons-dbcp & 0.71 & 0.95 & 0.82 & 0.71 & 0.64 & 0.90 & 0.95 & 0.92 & 0.90 & 0.73 & -11.4\\ 
		 commons-dbutils & 0.92 & 0.99 & 0.95 & 0.92 & 0.93 & 0.96 & 0.96 & 0.96 & 0.96 & 0.80 &  16.0\\ 
		 commons-digester & 0.74 & 0.80 & 0.77 & 0.74 & 0.81 & 0.73 & 0.87 & 0.79 & 0.73 & 0.73 & 10.6\\ 
		 commons-exec & 0.94 & 0.98 & 0.96 & 0.94 & 0.93 & 0.95 & 0.96 & 0.96 & 0.95 & 0.75 & 24.7\\ 
		 commons-fileupload & 0.92 & 0.93 & 0.93 & 0.92 & 0.95 & 0.88 & 0.89 & 0.89 & 0.88 & 0.91 & 3.7 \\ 
		 commons-io & 0.63 & 0.88 & 0.72 & 0.63 & 0.69 & 0.62 & 0.91 & 0.74 & 0.62 & 0.65 & 5.1 \\ 
		 commons-jelly & 0.87 & 0.96 & 0.91 & 0.87 & 0.92 & 0.72 & 0.76 & 0.74 & 0.72 & 0.82 & 12.4\\ 
		commons-jexl & 0.66 & 0.88 & 0.76 & 0.66 & 0.73 & 0.70 & 0.91 & 0.79 & 0.70 & 0.67 & 8.4 \\ 
		 commons-jxpath & 0.94 & 0.96 & 0.95 & 0.94 & 0.97 & 0.87 & 0.90 & 0.88 & 0.87 & 0.93 & 3.9\\ 
		 commons-net & 0.65 & 0.93 & 0.77 & 0.65 & 0.65 & 0.70 & 0.80 & 0.74 & 0.70 & 0.68  & -4.6\\ 
		 commons-validator & 0.93 & 0.96 & 0.94 & 0.93 & 0.95 & 0.94 & 0.95 & 0.94 & 0.94 & 0.95 & 0.0  \\ 
		 commons-vfs & 0.80 & 0.96 & 0.87 & 0.80 & 0.88 & 0.68 & 0.78 & 0.72 & 0.68 & 0.76 & 15.0\\ 
		 felix & 0.82 & 0.91 & 0.86 & 0.82 & 0.77 & 0.96 & 0.97 & 0.96 & 0.96 & 0.96 & -19.9 \\ 
		 httpcomponents-client & 0.89 & 0.93 & 0.91 & 0.89 & 0.94 & 1.00 & 1.00 & 1.00 & 1.00 & 1.00 & -5.8\\ 
		 httpcomponents-core & 0.94 & 0.97 & 0.95 & 0.94 & 0.97 & 0.94 & 0.96 & 0.95 & 0.94 & 0.97 & 0.0 \\ 
		 mina-sshd & 0.94 & 0.97 & 0.96 & 0.94 & 0.35 & 0.97 & 0.99 & 0.98 & 0.97 & 0.38 & -6.9\\ 
		 santuario-java & 0.79 & 0.89 & 0.84 & 0.79 & 0.87 & 0.78 & 0.90 & 0.84 & 0.78 & 0.85 & 2.2\\ 
		 zookeeper & 0.85 & 0.95 & 0.89 & 0.85 & 0.49 & 0.73 & 0.85 & 0.79 & 0.73 & 0.20 & 139\\ 
		\bottomrule
		\end{tabular}
\end{table*}

\section{Discussion}
\label{sec:discussion}
Our study evaluates the usefulness of tree kernels for clone and defect detection using two large datasets: BigCloneBench benchmark and Technical Debt dataset. Together these benchmarks provide a complete view of the tree kernels' ability for the evaluated tasks.
\subsection{Clone detection performance of tree kernels}
RQ1 to RQ3 experiments were conducted with known true and false clones from BigCloneBench, and every clone in the benchmark was used in the evaluation at least once, ensuring a realistic evaluation.
Out of the three tree kernels that we used, PTK did a better job overall, however, it also took more time and memory compared to the other tree kernels. 
It is anticipated since PTK matches partial trees, which are more in number than subtrees or subset trees. 
We also observed that just like any other clone detection method, our approach performs better for intra-project clone detection than inter-project. Similarly the high performance on Type-1 and Type-2 clones shows that it may be easier in practice to identify bugs in commits that are syntactically similar to the prior commits. Considering that a project might have more Type-3 clones than Type-1 and Type-2, the performance (on ST3 clones) is not remarkable. However, detecting Type-3 clones is still an open research problem and none of the existing tools are reported to do Type-3 clone detection reliably~\cite{roy2007survey}~\cite{sourcerercc}. 


\subsection{Defect detection performance of tree kernels}
Our main goal is to classify a new change by comparing it against historical changes. Prior work stresses the need to have a dataset with correct defect labels~\cite{herbold2019issues}, and therefore we use a publicly available dataset that was generated using a manually validated implementation of the SZZ algorithm~\cite{TDD}. The evaluation done on the Technical Debt dataset revealed a good Top-K accuracy for most projects. 
In our evaluation, Top-K drops to 0.6 for a few projects due to the limited number of commits available prior to the test commit at a particular point in time. If a project has sufficient data the Top-K is almost always higher than 0.6. Therefore, for the classification to be effective, a project has to have a huge corpus of historical changes, which is a practical limitation. 
\subsection{Implications}
\textbf{\textit{Usage scenario:}}
A typical usage scenario of our approach begins when a developer initiates a code commit. Before this commit reaches the central repository or branch, our approach analyzes the source code by comparing it against past changes in the project's version control system. If the new change matches an existing bug-inducing change, the developer is suggested to inspect their code. In this way, our approach prevents the bugs from reaching the central repository. Furthermore, reporting bugs at check-in time saves time and maintenance costs associated with fixing them at later stages.

\textbf{\textit{Defect granularity:}}
The large chunks of new source code introduced across changes in the Technical Debt dataset suggest that defect detection at commit level is not very useful. As for the line-level models, their results are also difficult to inspect because the reported bug-inducing lines cannot be analyzed in isolation, but rather, developers need additional context lines to understand bugs. The defects reported by our approach naturally contain the contextual information that the developers might need. Across the evaluated projects, the median method length is between $4$ and $14$. A developer, working on such a project will inspect only $4$ to $14$ lines to find the cause of bug.

\begin{figure}[h]
	\centering
	\includegraphics[width=1\columnwidth]{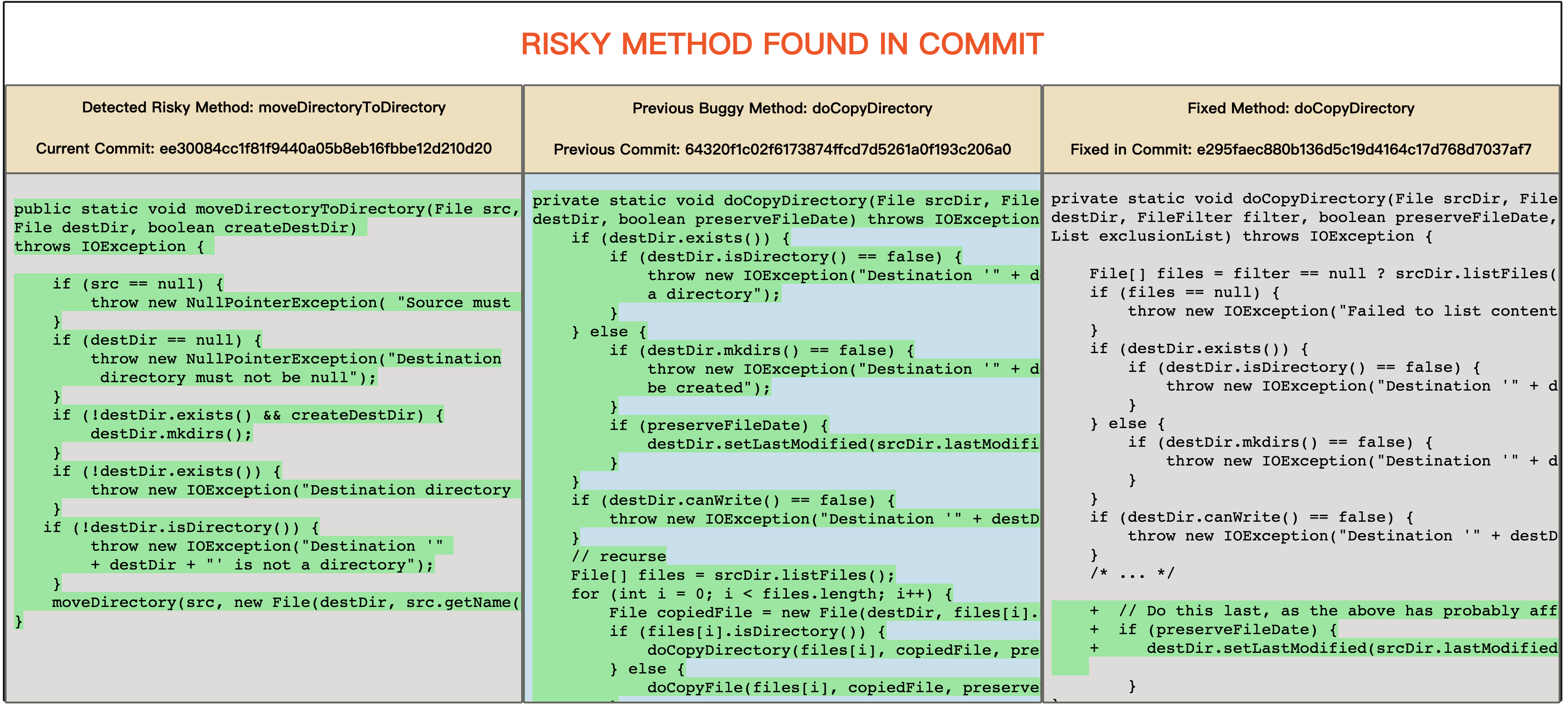}
	\caption{Example report of a risky commit from prototype.}
	\label{fig:tool}
\end{figure}

\textbf{\textit{Recommending fixes:}}
The output of the SZZ is pairs of bug-inducing and bug-fixing code, and the later might help in fixing new defects. 
After identifying a defect, our tool shown in Figure~\ref{fig:tool} proposes the fix associated with the past change that matched the new change. In future work, we plan to assess the usefulness of our approach for developers.

\subsection{Pros and Cons of tree kernels}
Tree kernels are computationally expensive, and our evaluation took many hours. The reason is a large number of changes and the need to compare each new change against all past changes, leading to $O(n^2)$ comparisons. 
However, classifying a new commit using tree kernels should be done within a minute, since changes from only one commit are compared against past commits. We further reduce the number of comparisons by using top 100 results from Elasticsearch, making the approach practical for real-world project development environments.
In our evaluation, we only assess the tree kernels on Java projects. Needless to say, the same
process can be developed with other programming languages simply by changing the language parser. 
This gives tree kernels an edge over language specific clone detectors. Lastly, tree kernels do not require fully parse-able code, hence, partial changes can be classified as long as their ASTs or similar trees can be generated.

\subsection{Threats to Validity}
We use an existing tree parser, Gumtree, and a tree kernels library, KeLP. While we check for errors, we cannot rule out that the \textit{internal validity} of our study is affected by relying on existing implementations. The performance of Elasticsearch affects and likely boosts the performance of our results, but it was applied to all tree kernels and baselines.
Our study uses a dataset generated by the SZZ algorithm which may have errors~\cite{herbold2019issues}, but the benchmark authors minimized this threat by manually validating their SZZ implementation following Kim \emph{et al.}~\cite{szz2006}, and Williams and Spacco~\cite{spacco2008szz}. 
Despite conducting evaluation on projects of different sizes and complexities, the \textit{external validity} is hampered by the use of only Apache foundation Java based projects.

\section{Conclusion}
Our paper is a first step in studying the effectiveness of tree kernels for clone detection and  JIT defect detection. To this end, we propose a \textit{simple} and \textit{explainable} tree kernels based approach that can classify a new method-level change as buggy or clean, and propose fixes to the developers. Our results suggest that of-the-shelf implementations of tree kernels are at least as effective as state-of-the-art clone detectors.
Therefore, future work should explore the untapped potential of tree kernels for different applications in our area. Motivated by these results, we plan to build and investigate an ensemble model in future work, by leveraging both clone detectors and tree kernels. 
As of now, our current approach may help practitioners to detect   bug-inducing changes, and equip them with the information they need to prioritize their inspection and testing efforts. \label{sec:conc}
\vspace{12pt}
\color{black}
\bibliography{mybibfile}
\end{document}